\documentclass[aps,prb,pra,twocolumn,floatfix,floats,showpacs,superscriptaddress]{revtex4}
\usepackage{amsmath, amsthm}
\usepackage{graphicx,latexsym}
\usepackage{dcolumn}
\newcommand{\ua}{\uparrow}
\newcommand{\da}{\downarrow}

\begin{document}

\title{Pure spin current generation in a Rashba-Dresselhaus quantum channel}

\author{Chia-Hui Lin}
\affiliation{Research Center for Applied Sciences, Academia Sinica,
Taipei 11529, Taiwan}

\author{Chi-Shung Tang}
\email[Email:\ ]{cstang@nuu.edu.tw}
\affiliation{Department of
Mechanical Engineering, National United University, Miaoli 36003,
Taiwan}

\author{Yia-Chung Chang}
\email[Email:\ ]{yiachang@gate.sinica.edu.tw} \affiliation{Research
Center for Applied Sciences, Academia Sinica, Taipei 11529, Taiwan}

\date{\today}

\begin{abstract}
    We demonstrate a spin pump to generate pure spin current of tunable
    intensity and polarization in the absence of charge current. The pumping functionality is
    achieved by means of an ac gate voltage that modulates the Rashba constant
    dynamically in a local region of a quantum channel with both static Rashba and Dresselhaus spin-orbit
    interactions. Spin-resolved Floquet scattering matrix is
    calculated to analyze the whole scattering process. Pumped spin current can be
    divided into spin-preserved transmission and spin-flip reflection
    parts. These two terms have opposite polarization of
    spin current and are competing with each other. Our proposed
    spin-based device can be utilized for non-magnetic control of spin
    flow by tuning the ac gate voltage and the driving frequency.
\end{abstract}

\maketitle

\section{Introduction}
Manipulation of electron spins can be achieved via applying external
active control, which is the essential requirement of spintronics
devices.\cite{spinintro}  Especially, spin-resolved current
generation is one of the key interests in spintronics research for
its potential application in quantum information
science.\cite{Zuti:0323,Burk:2007} Various approaches were proposed
to overcome the fundamental challenge in the issues of spin current
manipulation, detection, and injection efficiency. Methods based on
controlling magnetic field\cite{Wats:8301,Sun:8301, Zhan:6602} and
material ferromagnetism\cite{Brat:5304} are investigated. However,
for practical applications, more efficient methods that do not
involve strong magnetic field or interfaces between ferromagnets and
semiconductors are still needed. Spin pumping can be a viable
solution to the spin current generation.\cite{Shar:0531,Li:5312}

 Pumping of charge current is a fully quantum mechanical phenomenon
in a mesoscopic system that can generate current without applied
bias between two leads.  Theoretically and experimentally, charge
current pump has been realized and implemented in a quantum
channel or a cavity in the way of periodic
modulation.\cite{Thou:6083,Brou:0135,Swit:1905,Tang:353,Wang:3306}
In the adiabatic regime, Brouwer proposed a clear picture that the
pumped current depends on the enclosed area in parametric space
which is formed by a set of periodically varied parameters. Such
formalism was readily extended to non-adiabatic regime, which is
valid in the whole spectrum of frequency.\cite{Vavi:5313} If the
spin degree of freedom is incorporated, spin-dependent
transmission coefficients can be differentiated either directly by
external magnetic field\cite{Mucc:6802,Benj:5318} or by spin-orbit
interaction.\cite{Gove:5324,Li:5312} Spin pumping is generalized
from the quantum pumping and exempted from the spin injection
problem which occurs in the integrated semiconductor-ferromagnet
architecture.

In order to achieve spin pumping, a Rashba-type narrow channel
(which ignores the presence of the Dresselhaus term) driven by
local time-dependent potential was
proposed.\cite{Wang:5304,Tang:5804} When electrons propagate
through the potential region, quasi-bound state feature was shown
to enhance the spin-resolved transmission difference so that
sizable pure spin current can be generated. However, since the
Dresselhaus spin-orbit interaction is an intrinsic effect in
semiconductor materials with bulk inversion
asymmetry,\cite{Dresselhaus55} it is essential to take into
account this effort when considering such a spin pumping device.
It should be noted that the presence of the Dresselhaus term will
lead to the spin-flip mechanism which can modify the spin-pumping
characteristics in a qualitative way. We shall elucidate the
possibility to manipulate not only the intensity but also the
polarization of the spin current.

In this paper, the spin-resolved Floquet scattering matrix formalism
is applied to our system.\cite{Zhan:5347,Wu:5412} Based on the
Floquet theorem, this formalism provides an exact and
nonperturbative solution to the time-periodic Schr\"{o}dinger
equation in the mesoscopic system. Because the time-dependent
spin-orbit interaction couples two spin polarizations and all
sidebands together, analytic expression for the sideband dispersion
is not feasible. Thus, we determine the sideband dispersion relation
numerically by solving the Schr\"{o}dinger equation in a nearly
complete basis. Besides, the spatial inhomogeneity can also be
handled by matching boundary conditions piece by piece spatially.
The Floquet scattering matrix gives a coherent solution that goes
beyond the adiabatic regime.

\section{Model and Formalism}

\begin{figure}
    \centering
   \includegraphics[width=6cm]{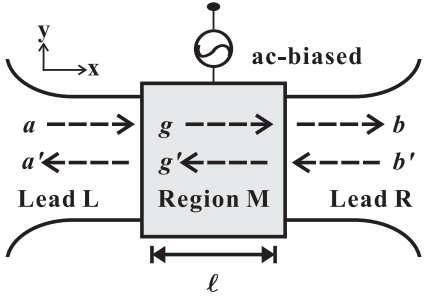}
  \caption{Schematic illustration of the quasi-1D spin-orbit quantum channel embedded in 2DEG. In this narrow channel,
  the electron gas has static Rashba and Dresselhaus spin-orbit interactions which are characterized by
  $\alpha_0$ and $\beta_0$ respectively. The central grey region,
  with width $l$, is biased by ac gate voltage so that Rashba strength locally modulated
  as $\alpha_1 \cos(\omega t)$. The origin of $x$-axis is set at the left edge of the grey region.}
  \label{fig1}
\end{figure}
    The system under consideration is a
two-dimensional electron gas (2DEG) that is present at the
interface of a heterostructure due to modulation doping and has
intrinsic static Rashba and Dresselhaus spin-orbit interactions.
The system configuration is shown in Fig. \ref{fig1}. A
quasi-one-dimensional (Q1D) narrow channel is formed from the 2DEG
via a lateral confining potential (along the $y$ direction).
The barrier separating the Q1D channel from the 2DEG should be strong
enough so the tunneling time between them is much longer than the
carrier transport time in the Q1D channel. A
finger gate is placed in the middle of the channel (the  grey
region in Fig. 1) that modulates the local Rashba interaction
strength sinusoidally via an ac-bias. Hence, the system can be
described by the effective Hamiltonian
\begin{equation}
\hat{\mathcal{H}} = \frac{\hat{p}^2}{2m^*}+\hat{\mathcal{H}}^{\rm
static}_{\mathrm{so}}+\hat{\mathcal{H}}_{\mathrm{so}}
(\mathbf{r},t)+ \hat{V}_c(y),
\end{equation}
where $m^*$ denotes the electron effective mass and
$\hat{V}_c(y)$ indicates the confinement potential in transverse
($y$) direction. $\hat{\mathcal{H}}^{\rm static}_{\mathrm{so}}$
and $\hat{\mathcal{H}}_{\mathrm{so}} (\mathbf{r},t)$ characterize,
respectively, the static and dynamic parts of spin-orbit
interaction. If we consider a narrow quantum channel where the
subband energy spacing is large enough to decouple $\hat{p}_y$
from spin-orbit interaction, the intersubband mixing is thus
neglected.\cite{Wang:5304,Tang:5804}  The longitudinal part of the
dimensionless Hamiltonian is then given by
\begin{eqnarray}
   \nonumber
  \hat{\mathcal{H}}^0_x &=& \hat{k}_x^2
- \alpha_0
 \sigma^y\hat{k}_x + \beta_0 \sigma^x \hat{k}_x,  \\
  \hat{\mathcal{H}}_x(t) &=& -\frac{1}{2}\alpha_1 \sigma^y \cos
(\omega t) \{\hat{k}_x,\theta(l/2-|x-l/2|)\},
 \label{Halmitonian}
\end{eqnarray}
where $\sigma^i(i=\{x,y,z\})$ denotes Pauli matrices and
$\hat{k}_x$ indicates the momentum operator $-i\partial_x$.
Anticommutator $\{\cdots\}$ is used to maintain the hermitianity
of $ \hat{\mathcal{H}}_x(t)$. The static Rashba strength
$\alpha_0$ is proportional the electric field perpendicular to the
interface where 2DEG lies. Additionally, $\beta_0$ is the
phenomenological Dresselhaus coupling parameter. In the finger
gate region, the Rashba parameter oscillates sinusoidally with
amplitude $\alpha_1$. For simplicity, we restrict the subsequent
discussions to the lowest subband and ignore the subband index.
The contributions from other subbands can be added if a more
realistic consideration is needed.

To proceed, it is convenient to rotate the spin quantization axis
such that  $\hat{\mathcal{H}}^0_x$ is diagonalized. The transformed
Hamiltonian is
\begin{eqnarray}
  \hat{\mathcal{H}}'^0_x    &=& \hat{k}_x^2- \gamma_0 \sigma^z \hat{k_x} ,  \label{H0}\\
  \hat{\mathcal{H}}'_x(t) &=& -\frac{1}{2}\alpha_1 \sigma^{\phi}  \cos (\omega
  t)\{\hat{k}_x,\theta(l/2-|x-l/2|)\} \label{H1},
\end{eqnarray}
where $\sigma^{\phi} = (\sigma^z \sin \phi  - \sigma^y\cos \phi  )$,
$\gamma_0 = \sqrt{\alpha^2_0+\beta^2_0}$, and
$\phi=\arctan(\alpha_0/\beta_0)$. $\hat{\mathcal{H}}'^0_x$
illustrates not only our choice of spin-up and spin-down states but
also that the location of subband bottom is at $-\gamma^2_0/4$.
Based on Floquet theorem, the wave functions in lead L ($x<0$) and
lead R ($x>l$) are given by
\begin{eqnarray}
\nonumber
  \Psi^{L}(x,t)&=& \sum_{m, \sigma}(a_{m,\sigma}e^{i k^R_{m,\sigma} x}
+a'_{m,\sigma}e^{i k^L_{m,\sigma} x} )e^{-i(\mu+m\omega )t}
\chi_{\sigma}, \\
\nonumber
   \Psi^{R}(x,t)&=& \sum_{m, \sigma}(b_{m,\sigma}e^{i k^R_{m,\sigma} x} +b'_{m,\sigma}e^{i k^L_{m,\sigma} x} )e^{-i(\mu+m\omega )t}
\chi_{\sigma},\\
\end{eqnarray}
where $\chi_{\sigma}$ denotes the spinor basis and $\mu$ represents
the incident energy. The sideband index $m$ runs essentially for all
integers.  From the dispersion relation in Eq.\ (\ref{H0}$),
k^R_{m,\sigma}$ and $k^L_{m,\sigma}$ are $\frac{1}{2}[\eta_{\sigma}
\gamma_0 + \sqrt{\gamma_0^2 + 4(\mu + m \omega)}]$ and
$\frac{1}{2}[\eta_{\sigma} \gamma_0 - \sqrt{\gamma_0^2 + 4(\mu + m
\omega)}]$ respectively, where $\eta_{\sigma}$ is defined as $
\sigma^z_{(\sigma,\sigma)}$. $a_{m,\sigma}$ ($a'_{m,\sigma}$) is the
amplitude of the rightward (leftward) wave in the \textit{m}th
sideband with spin $\sigma$ in lead L. Similarly, $b_{m,\sigma}$
($b'_{m,\sigma}$) is for lead R. Technically, these amplitudes are
determined by boundary condition and the direction of incident wave.

In the time-dependent region M ($0<x<l$), the general solution would
be
\begin{equation}\label{}
      \Psi^{M}(x,t) = \sum_{m, \sigma} \Psi_{m,\sigma}(x)e^{-i(\varepsilon +m\omega )t}
\chi_{\sigma},
\end{equation}
where $\varepsilon$ is the Floquet quasi-energy.
$\Psi_{m,\sigma}(x)$ is solved from Schr\"{o}dinger's equation,
\begin{eqnarray}
\nonumber
    \sum_{\sigma' } [(\hat{k}_x^2 - \gamma_0 \hat{k}_x \sigma^z_{(\sigma,\sigma')}) \Psi_{m,\sigma'}
    &-&
     \frac{\alpha_1 \hat{k}_x}{2} \sigma^{\phi}_{(\sigma, \sigma')} \\
     (\Psi_{m+1,\sigma'}  + \Psi_{m-1,\sigma'})] &=&  (\varepsilon+  m\omega)\Psi_{m,\sigma}.
\end{eqnarray}
These coupled equations can be expressed in matrix form,
\begin{equation}\label{quadeigen}
 \hat{k}_x^2 \mathbf{\Psi} + \hat{k}_x \mathbf{H^{(1)} \Psi} = \mathbf{H^{(0)} \Psi},
\end{equation}
where
\begin{eqnarray}
  \mathbf{H}^{\mathbf{(1)}}_{(m,\sigma)(m',\sigma')} &=& -\frac{\alpha_1}{2}\sigma^{\phi}_{(\sigma,
                                                    \sigma')}(\delta_{m, m'+1}+\delta_{m,
                                                    m'-1}) \nonumber \\
                                                    && - \gamma_0\sigma^z_{(\sigma ,
                                                    \sigma')} \delta_{m, m'} ,  \\
 \mathbf{H}^{\mathbf{(0)}}_{(m,\sigma)(m',\sigma')} &=& ( \varepsilon +m \omega)
                                                    \delta_{m, m'}\delta_{\sigma,
                                                    \sigma'}, \\
  \mathbf{\Psi}_{m,\sigma} &=& \Psi_{m,\sigma}(x).
\end{eqnarray}
Because this is a transport problem, we have to solve the
eigenvalue $q$ for fixed $\varepsilon$. This quadratic
eigenproblem can be solved by introducing another of auxiliary
equation $\mathbf{\Psi'} = q \mathbf{\Psi}$. Then Eq.
(\ref{quadeigen}) becomes
\begin{equation}
\begin{pmatrix}
  \mathbf{0}       & \mathbf{1}        \\
  \mathbf{H^{(0)}} & \mathbf{-H^{(1)}} \\
\end{pmatrix}
\begin{pmatrix}
 \mathbf{ \Psi }\\
 \mathbf{ \Psi'} \\
\end{pmatrix}
=q
\begin{pmatrix}
  \mathbf{\Psi}  \\
  \mathbf{\Psi'} \\
\end{pmatrix} .
\end{equation}
If we truncate the sideband index $m$ at $-M/2$ and $M/2$, where
$M$ is an even integer, the eigenvalues $q^j$ and eigenvectors
$\psi^j_{m,\sigma}$ are numerically determined from the above
secular equation.

Because Hamiltonian in Eq.\ (\ref{Halmitonian}) preserves
time-reversal symmetry, any $q^j$ is associated with $-(q^j)^*$,
i.e. $\varepsilon(q^j) = \varepsilon(-(q^j)^*)$. In addition, for
Hamiltonian is also invariant under inversion followed by spin
flip, $q^j$ has its another counterpart $-q^j$. Thus, we can
definitely sort the $(4M+4)$ complex eigenvalues into two groups.

For the case of evanescent modes, those right-decaying waves are
characterized by positive Im($q^j$); left-decaying waves have
negative Im($q^j$). On the other hand, for the case of propagating
modes that have real $q^j$, we sort $q^{j}$ with positive
(negative) group velocity to be rightward (leftward) propagating
waves. The group velocity is determined by
$(d\varepsilon/dq^j)$.\cite{Chan:0605} Therefore, the wave function
in region M is given by
\begin{equation}\label{}
    \Psi_{m,\sigma}(x) = \sum_{j} ( g_j \psi^{j,R}_{m, \sigma} e^{i
    q^{j,R}x} + g'_j \psi^{j,L}_{m, \sigma} e^{i q^{j,L}x} ),
\end{equation}
where superscripts $R$ and $L$ are added to indicate the propagating
or decaying direction.

Wave functions are matched in the time domain by $\varepsilon$ =
$\mu$ and continuous across the boundaries. Their derivatives
satisfy the following boundary conditions:
\begin{eqnarray}
\nonumber
  \partial_x \Psi(x,t)|_{x=0^+} -\partial_x \Psi(x,t)|_{x=0^-}&=&  \frac{i
  \alpha_1}{2}
\cos(\omega t)\sigma^{\phi} \Psi(0,t), \\
\nonumber
\partial_x \Psi(x,t)|_{x=b^-} -  \partial_x \Psi(x,t)|_{x=b^+}&=& \frac{ i \alpha_1}{2}  \cos(\omega
t)\sigma^{\phi}  \Psi(b,t).\\
\end{eqnarray}
The above boundary conditions can be written down in matrix form,
\begin{widetext}
\begin{eqnarray}
  \mathbf{a+a'} &=& \mathbf{S^R g+S^L g'}, \label{bc1}\\
  \mathbf{(K^R a+K^L a') -(S^RQ^Rg + S^LQ^Lg')} &=& \frac 1 2  \mathbf{(\mathbf{H^{(1)}}-\gamma_0 \mathbf{\Sigma})(a+a')},\\
  \mathbf{S^R}e^{i\mathbf{Q^R}l}\mathbf{g}+\mathbf{S^L}e^{i\mathbf{Q^L}l}\mathbf{g'} &=& e^{i\mathbf{K^R}l}\mathbf{b} +e^{i\mathbf{K^L}l}\mathbf{b'}, \\
  \mathbf{(K^R b+K^L b')}-(\mathbf{S^RQ^L}e^{i\mathbf{Q^R}l}\mathbf{g} +
\mathbf{S^LQ^L}e^{i\mathbf{Q^L}l} \mathbf{g'}) &=& \frac 1
2\mathbf{(\mathbf{H^{(1)}}-\gamma_0
\mathbf{\Sigma})}(e^{i\mathbf{K^R}l}\mathbf{b}
+e^{i\mathbf{K^L}l}\mathbf{b'})\label{bc4},
\end{eqnarray}
\end{widetext}
where those column vectors $\mathbf{a}$, $\mathbf{g}$, and
$\mathbf{b}$, are assigned values from amplitudes $a_{m,\sigma}$,
$g_j$, and $b_{m,\sigma}$ respectively. The above $(2M+2) \times
(2M+2)$ matrices $\mathbf{S}^{\mathbf{R}(\mathbf{L})}$,
$\mathbf{Q}^{\mathbf{R}(\mathbf{L})}$, $\mathbf{\Sigma}$, and
 $\mathbf{K}^{\mathbf{R}(\mathbf{L})}$ have matrix elements
\begin{eqnarray}
 \nonumber
  \mathbf{S}^{\mathbf{R}(\mathbf{L})}_{(m,\sigma),j} &=& \psi^{j,R(L)}_{m,\sigma}, \\
\nonumber
 \mathbf{Q}^{\mathbf{R}(\mathbf{L})}_{j,j'} &=& \delta_{j,j'}q^{j,R(L)}, \\
 \nonumber
  \mathbf{\Sigma}_{(m,\sigma)(m',\sigma')} &=& \delta_{m,m'} \delta_{\sigma,\sigma'}(-\delta_{\sigma, \ua} +
\delta_{\sigma, \da}),\\
\nonumber
  \mathbf{K}^{\mathbf{R}(\mathbf{L})}_{(m,\sigma)(m',\sigma')} &=& \delta_{m,m'} \delta_{\sigma,\sigma'}(\delta_{\sigma, \ua}
k^{R(L)}_{m,\ua} +\delta_{\sigma, \da}k^{R(L)}_{m,\da}). \\
\end{eqnarray}
After some algebra, we have the following matrix equation from Eqs.
(\ref{bc1}) to (\ref{bc4}):
\begin{equation}
    \begin{pmatrix}
  \mathbf{a'} \\
  \mathbf{b} \\
\end{pmatrix}
=  \begin{pmatrix}
  \mathcal{M}_{11}      & \mathcal{M}_{12}     \\
  \mathcal{M}_{21}  & \mathcal{M}_{22}  \\
\end{pmatrix}
\begin{pmatrix}
  \mathbf{a} \\
  \mathbf{b'} \\
\end{pmatrix}.
\label{SS}
\end{equation}
$\mathcal{M} =\begin{pmatrix}
  \mathcal{M}_{11}      & \mathcal{M}_{12}     \\
  \mathcal{M}_{21}  & \mathcal{M}_{22}  \\
\end{pmatrix}$ denotes $(4M+4) \times (4M+4)$ matrix
connecting the input coefficients with output coefficients including
all propagating and evanescent Floquet sidebands.

 In order to construct the Floquet scattering matrix, we need to introduce the concept
of probability flux amplitude into $\mathcal{M}$. We can
straightforward define a new matrix as
\begin{equation}
  \mathcal{M}'
=
\begin{pmatrix}
  \mathbf{V^L}      & 0    \\
  0 &   \mathbf{V^R}   \\
\end{pmatrix}
 \begin{pmatrix}
  \mathcal{M}_{11}      & \mathcal{M}_{12}     \\
  \mathcal{M}_{21}  & \mathcal{M}_{22}  \\
\end{pmatrix}
\begin{pmatrix}
  \mathbf{V^R}      & 0    \\
  0 &   \mathbf{V^L}   \\
\end{pmatrix}^{-1},
\end{equation}
where $\mathbf{V}^{\mathbf{R(L)}}_{(m,\sigma)(m',\sigma')}  =
 \delta_{m,m'} \delta_{\sigma,\sigma'}\sqrt{|2k^{R(L)}_{m,\sigma} - \eta_{\sigma}\gamma_0
 k^{R(L)}_{m,\sigma}|}$. In both leads, $\mathbf{V}^{\mathbf{R(L)}}$ takes the form
 of diagonal matrix with the square root of group velocity absolute value from each
 sideband and spin type. It is worth mention that $\mathcal{M}'$ is
 not unitary yet due to the presence of evanescent modes. In the
 final stage, we obtain a unitary Floquet scattering matrix by
 setting the evanescent modes of the total scattering matrix
 $\mathcal{M}'$ to be zero:
 \begin{equation}
  \mathcal{S} =  \begin{pmatrix}
  {\mathcal{R}}  & {\mathcal{T'} }    \\
  {\mathcal{T}}  & {\mathcal{R'}}  \\
\end{pmatrix}.
\end{equation}
The unitarity of Floquet scattering matrix reflects the current
conservation law,\cite{Li:5732,Hens:6218} and is used as the
criteria to check numerical convergence.

The reflection and transmission coefficients are readily obtained by
summing over matrix elements of $\mathcal{S}$. When electrons that
are incident from L lead with initial spin $\sigma_i$ are partially
reflected and transmitted to final spin $\sigma_f$, the
spin-resolved reflection and transmission coefficients are written
as
\begin{eqnarray}
  R_{\sigma_f \sigma_i}^{LL}(\varepsilon) &=& \sum_{m}|{\mathcal{R}}_{(m,\sigma_f)(0,\sigma_i)}|^2, \\
  T_{\sigma_f \sigma_i}^{RL}(\varepsilon) &=& \sum_{m}|{\mathcal{T}}_{(m,\sigma_f)(0,\sigma_i)}|^2.
\end{eqnarray}
On the contrary, if the electron is incident from lead R, this gives
rise to such reflection and transmission coefficients
\begin{eqnarray}
  R_{\sigma_f \sigma_i}^{RR}(\varepsilon) &=& \sum_{m}|{\mathcal{R'}}_{(m,\sigma_f)(0,\sigma_i)}|^2, \\
  T_{\sigma_f \sigma_i}^{LR}(\varepsilon) &=& \sum_{m}|{\mathcal{T'}}_{(m,\sigma_f)(0,\sigma_i)}|^2.
\end{eqnarray}
Under zero longitudinal bias, the spin-resolved current pumped out
through lead R is generally defined as
\begin{eqnarray}
\nonumber
    I^{R}_{\ua} &=& \frac{e}{h}\int d\varepsilon f(\varepsilon) [T_{\ua\ua}^{RL}+T_{\ua\da}^{RL}+R_{\ua\ua}^{RR}+R_{\ua\da}^{RR}-1],\\
\nonumber
    I^{R}_{\da} &=& \frac{e}{h}\int d\varepsilon f(\varepsilon)
    [T_{\da\ua}^{RL}+T_{\da\da}^{RL}+R_{\da\ua}^{RR}+R_{\da\da}^{RR}-1],\\
\end{eqnarray}
where $f(\varepsilon)$ is Fermi-Dirac distribution. The
spin-resolved current can be derived based on the framework of
B\"{u}ttiker's formula\cite{Butt:2485} by regarding two spin types
as different terminal channels. The generalization of
B\"{u}ttiker's formula for Floquet scattering matrix has been
strictly proven.\cite{Levi:1399,Kim:3309} The spin current and
charge current at lead R are defined as $I_{s}^{R} = I^{R}_{\ua} -
I^{R}_{\da}$ and $I_{c}^{R} = I^{R}_{\ua} + I^{R}_{\da}$. Because
system Hamiltonian in Eq.\ (\ref{Halmitonian}) has inversion
followed by spin flip symmetry, we can transform the transmission
and reflection coefficients as
$T_{\sigma_f\sigma_i}^{LR}=T_{-\sigma_f - \sigma_i}^{RL}$ and
$R_{\sigma_f \sigma_i}^{RR}=R_{-\sigma_f -\sigma_i}^{LL}$. Such
transformation firstly guaranteed that there is zero charge
current in this system. Secondly, when certain amount of spin
current is pumped out at lead R, there should be equal amount of
spin current with opposite polarization pumped out at lead L.
Furthermore, if this symmetry is combined with current
conservation condition, spin current formula can be simplified to
a more convenient form in calculation:
\begin{equation}
    I^R_s = \frac{2e}{h}\int d\varepsilon f(\varepsilon)
    [(T_{\ua\ua}^{RL}-T_{\da\da}^{RL})+(R_{\ua\da}^{RR}-R_{\da\ua}^{RR})].\\
\end{equation}
The first two terms represent contributions from transmitted
electrons, and the last two terms are attributed to reflected
electrons whose spin is changed. Hence, we separate $I^R_s$ into
spin-preserved transmission and spin-flip reflection parts because
their effects are different and discussed in the following context.
Thus,
\begin{equation}
    I^R_s = I^{R,{\rm trans}}_s+I^{R,{\rm refl}}_s .
\end{equation}
It should be noted that if there is no Dresselhaus term, the
$I^{R,{\rm refl}}_s$ term is identically zero, and the $I^R_s$ is
then reduced to the same form in Ref.\ \onlinecite{Wang:5304}.

\begin{figure}
    \centering
   \includegraphics[width=8.5cm]{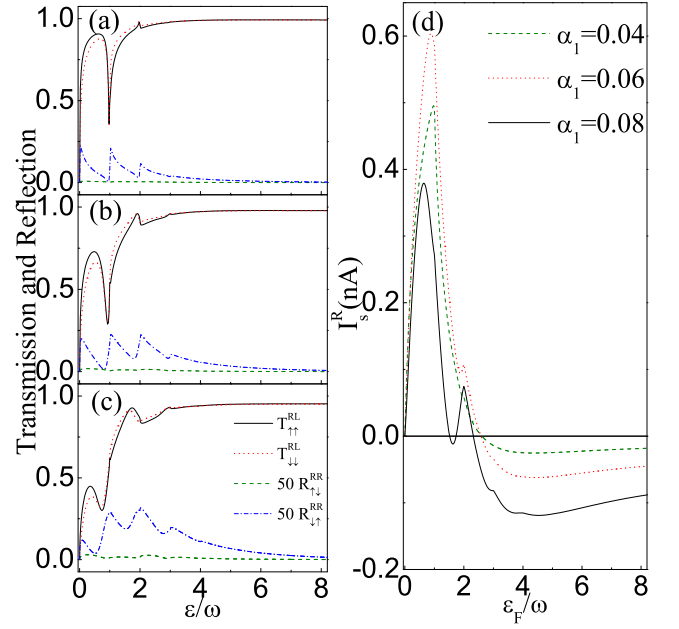}
  \caption{Spin-resolved transmission and reflection coefficients
  $T_{\ua\ua}^{RL}$, $T_{\da\da}^{RL}$, $R_{\ua\da}^{RR}$, and $T_{\da\ua}^{RR}$ as functions
  of the incident energy.  The values of reflection coefficients are
  multiplied by 50 to clarify the shape of the curves.
  $\alpha_0=0.12$, $\beta_0 =\alpha_0$, $l=30$, $\omega = 0.002$, and
  $\alpha_1=$(a) $0.04$, (b) $0.06$, (c) $0.08$. The spin current,
   $I_{s}^{R}$, which depend on Fermi energy for different $\alpha_1$ are plotted in (d).}
  \label{fig2}
\end{figure}
\begin{figure}
    \centering
   \includegraphics[width=8cm]{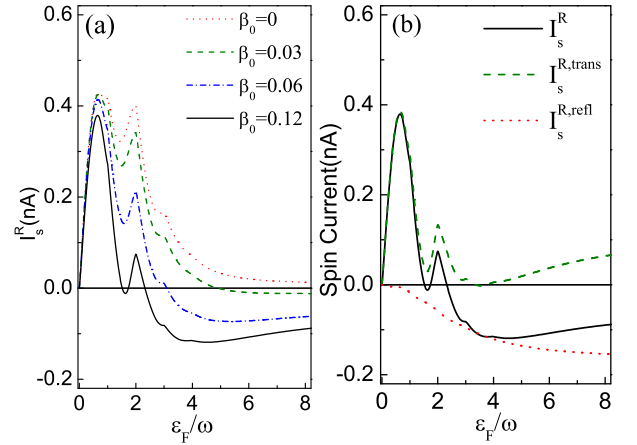}
  \caption{(a) Pumped spin current $I_{s}^{R}$ the Fermi energy (related to the bottom of the first subband in Q1D channel).
  $\beta_0$ is 0, 0.03, 0.06, and 0.12. Parameters $\alpha_1=0.08$,
  $\alpha_0=0.12$, $l=30$, and $\omega = 0.002$.
  (b)$I^R_s$, $I^{R,{\rm trans}}_s$, and $I^{R,{\rm refl}}_s$ are illustrated for $\beta_0=0.12$ case in (a). }
  \label{fig3}
\end{figure}

\section{Results and discussion}

Utilizing the above derived formula in previous section, it is easy
to calculate the spin current pumped from the spin-orbit quantum
channel via numerical means. The reasonable material parameters are
chosen from the narrow-gap heterostructure based on InGaAs-InAlAs
based system. According the experimental data, we assume that the
2DEG has an electron density $n_e =1\times10^{12}$ $\mathrm{cm}^{-2}$, effective
mass $m^*=0.04 m_0$, and $\alpha_0=0.12$ ($\hbar \alpha_0 =
2.8\times10^{-11}$ eV m).\cite{Nitt:1335}   The ratio between
Rashba and Dresselhaus terms can vary in certain range due to
experimental difficulties.\cite{Gigl:5327} Thus, we examine
the cases for $\beta_0/\alpha_0$ varying between 0 and 1.
In our calculations, the length and energy units are chosen to be $l^*=4.0$ nm
and $E^*=59$ meV (the Fermi energy of the 2DEG).
We assume that the ac-biased gate has a width of $l=30l^*$ and
its driving frequency is chosen as $\hbar \omega = 0.002 E^*$
($\omega/2\pi = 28$ GHz). The bottom of the lowest energy level (first subband) in the Q1D channel is
assumed to be slightly below the Fermi level, $E^*$ of the 2DEG so that the Fermi energy relative to the bottom
of the first subband in the Q1D channel (denoted $\varepsilon_F$) is comparable to $\hbar \omega$.
All numerical results are obtained for
zero temperature.

The dependence of transmission and reflection coefficients on the incident electron energy ($\varepsilon$)
for various values of $\alpha_1$ are illustrated in Figs.\ \ref{fig2}(a)-(c) when the static
Rashba and Dresselhaus constants are the same, i.e.
$\beta_0=\alpha_0$.  In order to clarify the important features shown in
these figures, we redefine the energy zero at the bottom of the first subband
with the presence of Rashba and Dresselhaus terms, i.e. $\hat{\mathcal{H}}'^0_x \rightarrow
\hat{\mathcal{H}}'^0_x +\gamma_0^2/4 $. The coefficients $T_{\ua\ua}^{RL}$,
$T_{\da\da}^{RL}$, $R_{\ua\da}^{RR}$, and $R_{\da\ua}^{RR}$, which are needed for calculating
$I^R_s$, are plotted in Figs. 2(a)-(c).  For transmission coefficients, we find sharp features at integer values of $\varepsilon/\omega$, indicative of the resonant inelastic
scattering.
As $\alpha_1$ increases, the dip around $\varepsilon/\omega=1$ moves toward
lower energy, and the dip width is broadened. The reason for the shift of dip
location is that a stronger oscillating potential would
lower the real part of the quasi-bound state energy and
shorten the lifetime of electrons trapped in such a
state.\cite{Li:5732} When $\alpha_1$ is increased to 0.08 as shown
in Fig. 2(c), a higher order resonance seen as a shallow dip around
$\varepsilon/\omega=2$ becomes more apparent because of the absorption
and emission of two quanta (with energy $2 \hbar \omega$). The most significant effect of
the Dresselhaus interaction is the emergence of the spin-flip process, which leads to
appreciable spin-flip reflection coefficients, $R_{\ua\da}^{RR}$ and $R_{\da\ua}^{RR}$.
In Figs. 2(a)-2(c), $R_{\ua\da}^{RR}$ and $R_{\da\ua}^{RR}$ have a saw-like
behavior with peaks appearing at integer values of $\varepsilon/\omega$,
where electrons are bounced back due to the presence of quasi-bound states.
Although their values are still minute compared with
$T_{\ua\ua}^{RL}$ and $T_{\da\da}^{RL}$, they can lead to significant change
in the final spin current when we take differences of the spin-up and spin-down contributions.

Figure 2(d) illustrates the spin current as a function of the Fermi energy, $\varepsilon_F$
(which reflects the carrier density in the Q1D channel) for various values of $\alpha_1$.
The curves in this Figs 2(a)-(c) can be
approximately divided into two parts: the low energy region
($\varepsilon/\omega<2$) and high energy region ($\varepsilon/\omega>2$). In the low
energy region, the reflection coefficients are too small compared to $(T_{\ua\ua}^{RL} -T_{\da\da}^{RL})$,
and $I^R_s$ is dominated by the contribution due to transmission process (denoted $I^{R,{\rm
trans}}_s$). In the high energy region, $(R_{\ua\da}^{RR}-R_{\da\ua}^{RR})$ becomes stronger than  $(T_{\ua\ua}^{RL} -T_{\da\da}^{RL})$ and the contribution to $I^R_s$ due to reflection process (denoted $I^{R,{\rm refl}}_s$) becomes dominant. As $\alpha_1$ increases
from 0.04 to 0.06, more spin current is pumped out
the first peak at $\varepsilon_F/\omega=1$ and into the second peak at $\varepsilon_F/\omega=2$. When $\alpha_1$ is tuned even higher to 0.08, high order
resonances become more relevant. Thus we have a further enhanced peak
around $\varepsilon_F/\omega=2$ and a reduced peak around $\varepsilon_F/\omega=1$.
However, because $(R_{\ua\da}^{RR}-R_{\da\ua}^{RR})$ is always
negative, $I^{R,{\rm refl}}_s$  results in negative contribution to
$I^R_s$ and it pulls the spin current curves downward. For $\varepsilon_F/\omega>2$, $I^{R,{\rm refl}}_s$ becomes dominant so that negative spin current is generated. As $\alpha_1$ increases, $I^R_s$ (for $\varepsilon_F/\omega>2$) becomes more negative due to higher probability of the spin-flip process.

In Fig.\ 3(a), we focus on the effect of Dresselhaus interaction on
the pumped spin current for a fixed $\alpha_1$.  In the case of zero
$\beta_0$, only one kind of spin polarization can be
pumped.\cite{Wang:5304}  As $\beta_0$ increases, $I^R_s$ curves tend
to shift downward due to increased spin-flip scattering process. In the
low density case ($\varepsilon_F/\omega<2$), experimentally reasonable $\beta_0$ may hardly
change the sign of $I_s^R$.  In the higher density case ($\varepsilon_F/\omega>2$), the sign of
$I^R_s$ is more vulnerable to the strength of the Dresselhaus term. When
$\beta_0$ is 0.03, 0.06, and 0.12, the threshold values of $\varepsilon$ at
which the sign of $I^R_s$ starts to change are at $\varepsilon_F/\omega$ =
4.89, 3.09, and 2.32, respectively.

A simple physical picture is presented here to give a conclusive
explanation.  The conditions in Fig.\ 3(b) are taken as an example.
Based on the dispersion relation of $\hat{\mathcal{H}}'^0_x$ in Eq.\
(\ref{H0}), when the electron is incident from lead L,
$|k^R_{0,\ua}|$ is always larger than $|k^L_{0,\da}|$ for the same
energy. Thus, it is easier for spin-up electron to tunnel through
this oscillating barrier due its larger flux, i.e. this dispersion
of static Hamiltonian tends to favor $T_{\ua\ua}^{RL}$ rather than
$T_{\da\da}^{RL}$. On the other hand, because scattering potential
$\hat{\mathcal{H}}'^1_x$ can be approximately regarded as
proportional to momentum, spin-up electrons could be more
susceptible to the scattering process so that $T_{\da\da}^{RL}$ is
favored here. In low energy region, these two mechanisms are
competing so that $(T_{\ua\ua}^{RL} -T_{\da\da}^{RL})$  may be
positive or negative and $I^{R,{\rm trans}}_s$ has obvious peaks.

In high energy region, because the second mechanism is less
relevant, only monotonically increasing $I^{R,{\rm trans}}_s$ is
present. For $I^{R,{\rm refl}}_s$, the situation is just on the
opposite side. Because $|k^L_{0,\da}|$ is greater than
$|k^L_{0,\ua}|$, there would be less chance for incident spin-down
electrons to be reflected. Hence, $I^{R,{\rm refl}}_s$ always
contributes to negative spin current and is monotonically
decreasing. When incident energy is low, $I^{R,{\rm refl}}_s$ only
compensates part of $I^{R,{\rm trans}}_s$. When energy increases,
$I^{R,{\rm refl}}_s$  becomes dominant and there is a threshold
$\varepsilon/\omega$ beyond which $I^{R}_s$ starts to change sign.

\section{conclusion}

We have proposed a promising approach to generate spin current non-magnetically
in the absence of charge current. A quasi-1D
channel with static Rashba and Dresselhaus spin orbit interaction is
studied. Spin pumping is achieved by an ac gate voltage to locally
modulate the Rashba constant. Pumped spin current can be attributed to
both the spin-preserved transmission and the spin-flip reflection processes. These two
terms contribute to opposite polarization of the spin current.

It is found that in the low density case ($\varepsilon_F/\omega<2$), the spin-preserved transmission is dominant and
featured by resonant inelastic scattering. In the high density case ($\varepsilon_F/\omega>2$),
there is a threshold  beyond which spin current begins
to switch polarization. Furthermore, it is found that the static
Dresselhaus coefficient $\beta_0$ as well as the dynamic Rashba
coefficient $\alpha_1$ can enhance the spin-flip process and modify
the threshold value of $\varepsilon_F/\omega$, at which the spin polarization switches. In conclusion, we have
demonstrated a feasible way to control dynamically the intensity
and polarization of the spin current via changing the strength of the
ac-biased gate voltage and tuning the driving frequency.

\begin{acknowledgments}
This work was supported in part by the National Science Council of
the Republic of China through Contract Nos. NSC95-2112-M-001-068-MY3
and NSC97-2112-M-239-003-MY3.
\end{acknowledgments}

\end{document}